\documentclass[onecolumn,showpacs,pra,10pt]{revtex4}
\usepackage{amsmath}
\usepackage{graphicx}

\input{tcilatex}

\begin{document}

\title{Bell-state preparation for fullerene based electron spins in distant
peapod nanotubes}
\author{W. L. Yang$,^{1,2}$ H. Wei$,^{1,2}$ X. L. Zhang$,^{3}$ M. Feng$^{1}$ 
}
\email{mangfeng@wipm.ac.cn}
\affiliation{$^{1}$State Key Laboratory of Magnetic Resonance and Atomic and Molecular
Physics, Wuhan Institute of Physics and Mathematics, Chinese Academy of
Sciences, Wuhan 430071, China }
\affiliation{$^{2}$Graduate School of the Chinese Academy of Sciences, Bejing 100049,
China}
\affiliation{$^{3}$Center for Modern Physics and Department of Physics, Chongqing
University, Chongqing 400044, China}

\begin{abstract}
We propose a potentially practical scheme, in combination with the
Bell-state analyzer [Zhang \textit{et al}., Phys. Rev. A \textbf{73}, 014301
(2006)], to generate Bell states for two electron spins confined,
respectively, in two distant $C_{60}$ fullerenes. To this end, we consider
the endohedral fullerenes staying in single walled carbon nanotubes ($SWCNTs$%
) and employ auxiliary mobile electrons and selective microwave pulses. The
application and the experimental feasibility of our scheme are discussed.
\end{abstract}

\maketitle

As a crucial resource of quantum information processing (QIP), entanglement
has exhibited peculiar correlation among the degrees of freedom of single
particles or the distinct parts of a composite system. Generally speaking,
creation of maximally entangled states between two qubits, i.e.,
Einstein-Podolsky-Rosen pairs or Bell states, is the first step towards more
complicated cases of entanglement. Motivated by potential applications of
entanglement, there are currently great interests in finding methods to
create and manipulate entangled states. Different schemes have been proposed
for realizing Bell states, and the relevant experimental demonstration has
been achieved in different systems \cite{review}.

One of the recently mentioned methods to generate entanglement is the use of
mobile qubits. As in \cite{moehring}, the trapped ions hold static qubits
encoded in atomic levels, and emit mobile qubits, i.e., photons, in a
controllable way. As the mobile qubits have entangled with the static
qubits, once we can entangle and then detect the mobile qubits, we will have
the entanglement of the static qubits. In this Brief Report, we will try to
move such an idea to an entanglement generation of two distant
fullerene-based electron spins. Doped fullerenes, like $^{15}N@C_{60}$ or $%
^{31}P@C_{60}$, have been considered as excellent candidates for spin-based
QIP \cite{Harneit, Suter, Twamley, Feng1}, the most attractive feature of
which is the long decoherence time of the electron spin of the doped atom
due to the protection from the fullerene cage. In the original QIP schemes
with endohedral fullerenes, the qubits could be encoded in either electron
spins \cite{Harneit} or nuclear spins \cite{Suter} of the doped atoms. No
matter which qubits are employed, the two-qubit gating is based on the
dipole-dipole coupling of the electron spins between nearest-neighbor
fullerenes. As a result, to entangle two distant qubits, we have to involve
a lot of overhead. In this work, we consider an efficient entanglement
scheme for distant fullerene-based qubits confined in single-walled carbon
nanotubes ($SWCNTs$) \cite{Iijima, Bock, Bena, Smith, Gunlycke}. The key
idea is that we inject auxiliary single electrons with certain polarization
to the potential wells in $SWCNTs$. After entangling the mobile qubits
(i.e., the spins of the injected electrons) with the static qubits (i.e.,
the electron spins inside the fullerenes), we move the mobile qubits away
and make them entangled with each other by a Bell-state analyzer \cite{Bee,
Zhang}. Then the static qubits, no matter how distant they are, will be
entangled in one of the Bell states.

$SWCNTs$ carry promises to transport information from one location to
another. As shown in \cite{Bock, Bena}, the quasi-one-dimensional nanosized
structure of the $SWCNTs$ offers the possibility of moving qubits in a
solid-state system. Experimentally, the transport of spin-polarized
electrons in a carbon nanotube has already been achieved \cite{Tsuk} and
well investigated \cite{Bena, Frank}. Moreover, with current technology, it
is possible to trap empty or doped fullerenes in hollow $SWCNTs$, which are
called fullerene peapods \cite{Chadli}. By using two fullerene peapods, we
will demonstrate below how to entangle two distant encapsulated electron
spins.

We sketch our scheme in Fig. 1, which consists of two distant $%
^{15}N@C_{60}@SWCNTs$ peapods encapsulating the static qubits $A^{^{\prime
}} $ and $B^{^{\prime }}$, respectively. Using turnstile injectors, we could
have two auxiliary electrons, carrying mobile qubits $A$ and $B$, injected
and confined respectively in shallow potential wells in the conduction bands
of the $SWCNTs$. We suppose each potential well to be shallow enough to hold
only a single electron. As the electron spin of the doped atom $^{15}N$ is $%
3/2,$ we encode the static qubits in the Zeeman levels $\left\vert
-3/2\right\rangle =\left\vert \uparrow \right\rangle _{A^{\prime }(B^{\prime
})}$ and $\left\vert 3/2\right\rangle =\left\vert \downarrow \right\rangle
_{A^{\prime }(B^{\prime })}.$ The mobile spins $A$ and $B$ are defined as $%
\left\vert -1/2\right\rangle =\left\vert \uparrow \right\rangle _{A(B)}$ and 
$\left\vert 1/2\right\rangle =\left\vert \downarrow \right\rangle _{A(B)}.$
Let us first consider the entanglement between the mobile qubit $A$ and the
static qubit $A^{^{\prime }}$. Under magnetic field gradient, $A$ and $%
A^{^{\prime }}$ with different level splitting, interact by dipole-dipole
coupling. Neglecting the negligible terms associated with nuclear spins, we
have the Hamiltonian in units of $\hbar =1,$%
\begin{equation}
H=g\mu _{B}B_{A}S_{z}^{A}+g\mu _{B}B_{A^{^{\prime }}}S_{z}^{A^{\prime
}}+JS_{z}^{A}S_{z}^{A^{\prime }},
\end{equation}%
where $\mu _{B}$ is the Bohr magneton, $g$ is the electron Land\'{e}
g-factor, and $B_{i}$ ($i=A$ and $A^{^{\prime }}$) is the magnetic field
strength experienced by qubit $i$. As mentioned in \cite{Suter},
individually addressing of qubits is available with selective microwave
pulses under the magnetic field gradients generated by micropatterned wires,
which can shift the resonance frequency between neighboring electron spins
by several Megahertz. In Eq.(1), the magnetic dipolar coupling strength $J$
between qubits $A^{^{\prime }}$ and $A$ could be expressed as $%
J=J_{0}(1-3\cos ^{2}\phi )$ \cite{Harneit}$,$ where $J_{0}=\hbar \gamma
^{2}/|r|^{3}$ with $\gamma $ the gyromagnetic ratio of the electron, $r$ the
distance vector between the two electron spins, and $\phi $ the angle
between $r$ and the magnetic field. In our case, $J$ is estimated to be $50$
MHz provided $|r|=1.14$ nm and $\phi =0$ as done in \cite{Suter}.
Furthermore, for two nearest-neighbor electron qubits distant by $1.14$ nm
in the magnetic field gradient $dB/dz=4\times 10^{5}$ T/m, the differences
of the electron spin resonance (ESR) frequencies between neighboring
encapsulated spins are about $12.7$ MHz and $12.7\times 3\approx 38$ MHz
regarding $\left\vert \pm 1/2\right\rangle $ and $\left\vert \pm
3/2\right\rangle $, respectively, and thereby the single-qubit operation is
available using narrow-band ESR pulses \cite{Feng1}.

Straightforward calculations show that the eigenenergies of Eq. (1) are%
\begin{eqnarray}
&&2\omega _{1}+3\omega _{2}+3J/2,\text{ \ \ \ }2\omega _{1}+\omega _{2}+J/2,%
\text{ \ \ \ \ \ }2\omega _{1}-\omega _{2}-J/2,\text{ \ \ \ \ }2\omega
_{1}-3\omega _{2}-3J/2,  \notag \\
&&-2\omega _{1}+3\omega _{2}-3J/2,\text{ }-2\omega _{1}+\omega _{2}-J/2,%
\text{ }-2\omega _{1}-\omega _{2}+J/2,\text{ }-2\omega _{1}-3\omega
_{2}+3J/2,
\end{eqnarray}%
in the subspace spanned by\bigskip\ $\left\vert \frac{1}{2},\frac{3}{2}%
\right\rangle ,$ $\left\vert \frac{1}{2},\frac{1}{2}\right\rangle ,$ $%
\left\vert \frac{1}{2},-\frac{1}{2}\right\rangle ,$\ $\left\vert \frac{1}{2}%
,-\frac{3}{2}\right\rangle ,$ $\left\vert -\frac{1}{2},\frac{3}{2}%
\right\rangle ,$ $\left\vert -\frac{1}{2},\frac{1}{2}\right\rangle ,$ $%
\left\vert -\frac{1}{2},-\frac{1}{2}\right\rangle ,$ $\left\vert -\frac{1}{2}%
,-\frac{3}{2}\right\rangle ,$ where $\omega _{1}=g\mu _{B}B_{A}/2$, and $%
\omega _{2}=g\mu _{B}B_{A^{^{\prime }}}/2$, and $S_{z}^{A}$ and $%
S_{z}^{A^{\prime }}$ in Eq. (1) are denoted by $S_{z}^{A}=%
\begin{pmatrix}
1 & 0 \\ 
0 & -1%
\end{pmatrix}%
\otimes 
\begin{pmatrix}
1 & 0 & 0 & 0 \\ 
0 & 1 & 0 & 0 \\ 
0 & 0 & 1 & 0 \\ 
0 & 0 & 0 & 1%
\end{pmatrix}%
,$ \ $S_{z}^{A^{\prime }}=%
\begin{pmatrix}
1 & 0 \\ 
0 & 1%
\end{pmatrix}%
\otimes \frac{1}{2}%
\begin{pmatrix}
3 & 0 & 0 & 0 \\ 
0 & 1 & 0 & 0 \\ 
0 & 0 & -1 & 0 \\ 
0 & 0 & 0 & -3%
\end{pmatrix}%
.$As depicted in Fig. 2, the splitting of a spin state is heavily dependent
on another coupled spin state. So some of the degeneracy are released. As a
result, a $\pi /2$ ESR pulse with the frequency $2\omega _{2}+J$ flip only
the target state of the static qubit $A^{^{\prime }}$ (i.e., $\left\vert
\uparrow \right\rangle _{A^{\prime }}\rightleftharpoons \left\vert
\downarrow \right\rangle _{A^{\prime }}$) in the case of the control state
of the mobile qubit $A$ being $\left\vert 1/2\right\rangle _{A}$ \cite{Feng1}%
. This is a nontrivial two-qubit gate $CNOT_{AA^{^{\prime }}}$. Similarly,
we can also construct another indispensable two-qubit gates $%
CNOT_{BB^{^{\prime }}}$ to entangle the mobile and static qubits.

Suppose that the mobile spin is initially prepared in a\ superposition state 
$(\left\vert \uparrow \right\rangle _{A(B)}+\left\vert \downarrow
\right\rangle _{A(B)})/\sqrt{2},$ and the state of the static qubit is $%
\left\vert \downarrow \right\rangle _{A^{\prime }(B^{\prime })}.$\ From the
initial state $\left\vert \Psi _{0}\right\rangle =(\left\vert \uparrow
\right\rangle _{A}\left\vert \downarrow \right\rangle _{A^{^{\prime
}}}+\left\vert \downarrow \right\rangle _{A}\left\vert \downarrow
\right\rangle _{A^{^{\prime }}})\otimes (\left\vert \uparrow \right\rangle
_{B}\left\vert \downarrow \right\rangle _{B^{^{\prime }}}+\left\vert
\downarrow \right\rangle _{B}\left\vert \downarrow \right\rangle
_{B^{^{\prime }}})/2,$ the gating $CNOT_{AA^{^{\prime }}}$ and $%
CNOT_{BB^{^{\prime }}}$ lead to%
\begin{equation}
\left\vert \Psi _{0}\right\rangle ^{^{\prime }}=\frac{1}{2}(\left\vert
\uparrow \right\rangle _{A}\left\vert \downarrow \right\rangle _{A^{^{\prime
}}}+\left\vert \downarrow \right\rangle _{A}\left\vert \uparrow
\right\rangle _{A^{^{\prime }}})\otimes (\left\vert \uparrow \right\rangle
_{B}\left\vert \downarrow \right\rangle _{B^{^{\prime }}}+\left\vert
\downarrow \right\rangle _{B}\left\vert \uparrow \right\rangle _{B^{^{\prime
}}}).
\end{equation}

Then the mobile qubits will be moved to the Bell-state analyzer\ designed in
Refs. \cite{Bee, Zhang} which makes use of the commutability between the
spin and charge degrees of freedom of the electron. As plotted in Fig. 1,
the two mobile qubits are sent through the analyzer from the ports \textit{a}
and \textit{b}. By making measurement by charge detectors, as listed and
explained in Table I, we could have the two mobile qubits entangled in a
specific Bell state$,$ which also implies a specific entanglement between
the two static qubits.

However, there was no imperfection considered in above treatment. Recent
observation \cite{Simon} has shown that $T_{1}$ of the doped electron spin
in $SWCNTs$ is $13$ $\mu s$ at $300$ K$,$ and $30$ $\mu s$ at $5$ K,
respectively, which are shorter compared to the crystalline cases \cite%
{Knorr}. It was speculated that the short $T_{1}$ is due to interaction of
the encapsulated electron spin with the nuclear spins in the host $SWCNTs$ 
\cite{Simon}. However, as the cage could reduce the detrimental effect from
decoherence to 25\% \cite{Twamley}, we may suppose below that the mobile
qubits are affected by decoherence more than the static qubits inside the
cages by four times, and thereby we would only consider the decaying of the
mobile qubits in our following treatment. Moreover, dephasing in our case is
strongly related to the external magnetic field. As the static qubit is
initially in the well polarized state, but the mobile qubit in superposition
state, we will only focus our attention on the dephasing of the mobile
qubits \cite{explain3}. Supposing our implementation is fast enough so that
no spin flip has actually happened, we consider following effective
Hamiltonian $%
%
H_{D}^{j}=\theta \sigma _{j}^{z}-i\frac{\Gamma _{1}^{j}}{2}\sigma _{j}^{+}\sigma _{j}^{-},%
$ where $\sigma _{j}^{k}$ $(k=z,+,-)$ are Pauli operators for the mobile
qubits with $j=A,$ $B,$ $\Gamma _{1}^{j}$ is the spin-flip relaxation rate
and $\theta $ is the level splitting plus the level shift due to coupling to
the static qubit. From the initial state $\left\vert \Psi \right\rangle
_{0}^{j}=(\left\vert \uparrow \right\rangle _{j}+\left\vert \downarrow
\right\rangle _{j})/\sqrt{2},$ we have the time evolution yielding $%
\left\vert \Psi (t)\right\rangle ^{j}=[(\cos (\theta t)-i\sin (\theta t))e^{-%
\frac{\Gamma _{1}^{j}}{2}t}\left\vert \uparrow \right\rangle _{j}+(\cos
(\theta t)+i\sin (\theta t))\left\vert \downarrow \right\rangle _{j}]/\sqrt{2%
}.$ To eliminate the dephasing effect, we could employ a trick by setting
the gating time to be $t_{g}=2k\pi /\theta ,$ with $k$ a constant determined
later. So we have $\left\vert \Psi (t)\right\rangle ^{j}=(e^{-\frac{\Gamma
_{1}^{j}}{2}t}\left\vert \uparrow \right\rangle _{j}+\left\vert \downarrow
\right\rangle _{j})/\sqrt{2},$ which is only suffered from the spin-flip
errors. To perform the $CNOT_{AA^{\prime }}$ or $CNOT_{BB^{\prime }}$ with
this trick, we should have $k=\theta /(2\Omega _{e})$ with $\Omega _{e}$ the
Rabi frequency under the radiation of ESR pulses. As a result, the entangled
state produced would not be affected by dephasing errors. To visualize the
effect of spin-flip relaxation, we have plotted in Fig. 3 the fidelity of
the generated Bell states $\left\vert \Psi \right\rangle _{A^{^{\prime
}}B^{^{\prime }}}^{\pm }$ and $\left\vert \Phi \right\rangle _{A^{^{\prime
}}B^{^{\prime }}}^{\pm }$\ in the dissipative situation, which are,
respectively, $F_{\left\vert \Psi \right\rangle _{A^{^{\prime }}B^{^{\prime
}}}^{\pm }}=(\alpha _{A}+\alpha _{B})^{2}/(1+\alpha _{A}^{2})(1+\alpha
_{B}^{2})$ and $F_{\left\vert \Phi \right\rangle _{A^{^{\prime }}B^{^{\prime
}}}^{\pm }}=(1+\alpha _{A}\alpha _{B})^{2}/(1+\alpha _{A}^{2})(1+\alpha
_{B}^{2}),$ with $\alpha _{A}=e^{-\Gamma _{1}^{A}t}$ and $\alpha
_{B}=e^{-\Gamma _{1}^{B}t}.$ Although the fidelities are generally going
down with increasing relaxation rates, the Bell state $\left\vert \Phi
\right\rangle _{A^{^{\prime }}B^{^{\prime }}}^{\pm }$ always keeps unit in
the case of $\Gamma _{1}^{A}=\Gamma _{1}^{B},$ which could be explained by
above expression $F_{\left\vert \Phi \right\rangle _{A^{^{\prime
}}B^{^{\prime }}}^{\pm }}$ and is reflected in the inset of Fig. 3.

We address some remarks for experimental implementation of our scheme. First
of all, in above treatment for imperfection, we have only considered the
dephasing due to external magnetic field, which is avoidable by our trick.
However, the intrinsic dephasing errors due to the nuclear spin of the
impurity atom $^{13}C$ is hard to be overcome. In the absence of external
magnetic field, $T_{2}$ remains about 20 $\mu s$ at 5 K for both fullerene
peapods and crystalline fullerenes \cite{Simon}. As $T_{2}$ is shorter than $%
T_{1},$ we should consider it seriously in designing our scheme. Our
operations include transport of mobile qubits and logic gating. It was
reported \cite{Bena, Frank} that the transport time of the mobile qubit with
Fermi velocity $10^{6}$ m/s could be as short as 1 picosec over a 1 $\mu m$ $%
SWCNT$ \cite{explain}, and our operation time shown above could be shorter
than 0.1 $\mu s$ in the case of $\Omega _{e}=25$ MHz$.$ As a result, the
influence from intrinsic dephasing is negligible with respect to our
implementation time. Secondly, we have neglected $SWCNT-SWCNT$ interaction 
\cite{Popescu}\ and $C_{60}-SWCNT$ interaction \cite{Okada, Dubay}, which
are far from the resonance frequency of the electron spin under our
consideration. In experiments, however, the $C_{60}-SWCNT$ interaction,
although very weak, should be seriously considered, which would influence
the transport of the mobile electron. Thirdly, no operational imperfection
was involved in our treatment above. Actually any deviation from the desired
time in switching off the potential well and in radiating the ESR pulse
would yield additional phases and lower the fidelity. But a nearly perfect
operation of our scheme seems challenging with current technology.

Besides the application mentioned in Fig. 1, our idea could also be
generalized to entangling two or many spatially separated static qubits in
the same $SWCNT$ (See Fig. 4) and to fusing two entangled states prepared
respectively in two $SWCNTs$ \cite{explain2}. As shown in Fig. 4(a), the
output electron with different spin polarization from a $SWCNT$ goes in
different ways due to the polarizing beam splitter, and then gets recorded
by the charge detector. The Bell states of the two static qubits $\left\vert
\Phi \right\rangle ^{+}=(\left\vert \downarrow \right\rangle \left\vert
\downarrow \right\rangle +\left\vert \uparrow \right\rangle \left\vert
\uparrow \right\rangle )/\sqrt{2}$\ and $\left\vert \Phi \right\rangle
^{-}=(\left\vert \downarrow \right\rangle \left\vert \downarrow
\right\rangle -\left\vert \uparrow \right\rangle \left\vert \uparrow
\right\rangle )/\sqrt{2}$\ are thereby generated in the case of $P_{e}=0$
and $P_{e}=1,$ respectively. Likewise, the GHZ state of $n$ static qubits $%
\left\vert GHZ\right\rangle ^{\pm }=(\left\vert \downarrow \right\rangle
_{1}\left\vert \downarrow \right\rangle _{2}\cdot \cdot \cdot \left\vert
\downarrow \right\rangle _{n}\pm \left\vert \uparrow \right\rangle
_{1}\left\vert \uparrow \right\rangle _{2}\cdot \cdot \cdot \left\vert
\uparrow \right\rangle _{n})/\sqrt{2}$ could also be produced when $P_{e}=0$
or $P_{e}=1$ (See Fig. 4(b)). Comparing to a previous work with two
electrons interacting and entangling in a $SWCNT$ \cite{Gunlycke}, our
scheme is relatively simpler. As the wavefunction of the static electron is
completely compressed in the cage, there is no wavefunction overlap between
the static and mobile electrons in our case. As a result, no concern about
the quantum characteristic of two identical particles is needed in our case,
and the entanglement between the mobile and static qubits could be
deterministically achieved by magnetic dipole-dipole interaction.

In summary, we have proposed a potential scheme to entangle two electron
spins in distant fullerene peapods. To accomplish our idea, we have to
employ ESR selective pulses, magnetic field gradient, and $SWCNTs$ with
switchable potential wells. Although some of the operations in our scheme
are still unreachable with current techniques, we argue that our scheme
would be helpful for achieving large-scale QIP setup with fullerene-based
qubits.

This work is supported by NNSF of China under Grants No. 10774163 and No.
10774042.

FIG. 1 Schematic setup to generate entanglement between two electron spins
in two distant endohedral fullerene peapods in $x-y$ plane under magnetic
field gradient $B_{Z}$. The static qubits $A^{^{\prime }}$ and $B^{^{\prime
}}$ are caged in fullerenes inside the $SWCNTs$. The mobile qubits $A$ and $%
B $ are injected and bound in shallow potential wells, which could be
switched on and off at will. The mobile qubits, after entangling with the
static qubits by ESR pulses (denoted by wavy arrows), will be moved to the
Bell-state analyzer in the dashed box (See Fig. 3(a) in \cite{Zhang}) where $%
P_{1}$ and $P_{2}$ are encoders which perform the spin parity measurement
and record bunching ($P_{1}=P_{2}=0$) and antibunching ($P_{1}=P_{2}=1$). $H$
means a Hadamard gate $H=(\sigma _{x}+\sigma _{y})/\sqrt{2}.$

FIG. 2 The eigenenergy spectrum of the two electron spins coupled by the
magnetic dipole-dipole interaction, where $|\cdot \cdot \rangle $ is the
state consisting of a mobile and a static qubits, and \{$\cdot \cdot \cdot $%
\} at the bottom and on the right represent, respectively, the degenerate
frequency difference between the nearest-neighbor levels along a column and
along a line.

FIG. 3 The fidelity of the Bell states $\left\vert \Psi \right\rangle
_{A^{^{\prime }}B^{^{\prime }}}^{\pm }$ (bottom surface) and $\left\vert
\Phi \right\rangle _{A^{^{\prime }}B^{^{\prime }}}^{\pm }$ (top surface) in
a dissipative situation$,$ where $K_{1}=\Omega _{e}/\Gamma _{1}^{A}$ and $%
K_{2}=\Omega _{e}/\Gamma _{1}^{B}.$ The $CNOT$ gating time we set is $%
t_{g}=\pi /\Omega _{e}$ with $\Omega _{e}=25$ MHz. The inset shows the
fidelity of the Bell states in the case of $K_{1}=K_{2}=K,$\ where the solid
and the dashed curves represent the fidelity of the Bell states $\left\vert
\Psi \right\rangle _{A^{^{\prime }}B^{^{\prime }}}^{\pm }$ and $\left\vert
\Phi \right\rangle _{A^{^{\prime }}B^{^{\prime }}}^{\pm },$ respectively.

FIG. 4 The schematics for an application of our scheme, where the mobile
qubit is initially prepared in a\ superposition state $(\left\vert \uparrow
\right\rangle +\left\vert \downarrow \right\rangle )/\sqrt{2},$ and a line
of endohedral fullerenes with static qubits prepared in $\left\vert
\downarrow \right\rangle _{1}\left\vert \downarrow \right\rangle _{2}\cdot
\cdot \cdot \left\vert \downarrow \right\rangle _{n}$ are confined in a $%
SWCNT$. By employing bias voltages and ESR pulses, in combintion with the
Hadamard gate, the charge detector $P_{e}$ (counting the charge number 0, 1
and 2 \cite{Bee}) and a polarizing beam splitter (denoted by double dashed
lines) which transmits spin up and refelcts spin down, we could creat
entanglement of the static qubits in a controllable way. (a) The creation of
Bell states of two spatially separated static qubits; (b) The creation of
GHZ states of $n$ static qubits.

\bigskip\ 

TABLE I. List of the resulting Bell states corresponding to different
outputs from the encoders $P_{1}$ and $P_{2}$. From

\cite{Zhang}, the entangled state of the mobile electrons A and B, input
respectively from port $a$ and $b,$ could be

$\left\vert \Psi \right\rangle _{AB}^{\pm }$ if $P_{1}=0$ or $\left\vert
\Phi \right\rangle _{AB}^{\pm }$ if $P_{1}=1.$ After the Hadamard gating, $%
P_{2}=1$ may correspond to $\left\vert \Psi \right\rangle _{AB}^{+}$ or $%
\left\vert \Phi \right\rangle _{AB}^{+}$

and $P_{2}=0$ means $\left\vert \Psi \right\rangle _{AB}^{-}$ or $\left\vert
\Phi \right\rangle _{AB}^{-}.$ Therefore, considering the detection results
from $P_{1}$ and $P_{2}$ together$,$ we could

specify the entanglement between the mobile qubits A and B, which yields the
entanglement between A' and B' by Eq. (3).

\ \ \ \ \ \ \ \ 
\begin{tabular}{l|l|l}
\hline\hline
\ Outputs from encoders & $\ \ \ \ \ \ \ \ \ \ \ \ \ \ \left\vert \Psi
\right\rangle _{AB}$ & $\ \ \ \ \ \ \ \ \ \ \ \ \ \ \left\vert \Psi
\right\rangle _{A^{^{\prime }}B^{^{\prime }}}$ \\ \hline
$\ \ P_{1}=0;$ $P_{2}=1$ & $\left\vert \Psi \right\rangle _{AB}^{+}=\tfrac{1%
}{\sqrt{2}}(\left\vert \uparrow \right\rangle _{A}\left\vert \downarrow
\right\rangle _{B}+\left\vert \downarrow \right\rangle _{A}\left\vert
\uparrow \right\rangle _{B})$ & $\left\vert \Psi \right\rangle _{A^{^{\prime
}}B^{^{\prime }}}^{+}=\frac{1}{\sqrt{2}}(\left\vert \downarrow \right\rangle
_{A^{^{\prime }}}\left\vert \uparrow \right\rangle _{B^{^{\prime
}}}+\left\vert \uparrow \right\rangle _{A^{^{\prime }}}\left\vert \downarrow
\right\rangle _{B^{^{\prime }}})$ \\ 
$\ \ P_{1}=0;$ $P_{2}=0$ & $\left\vert \Psi \right\rangle _{AB}^{-}=\frac{1}{%
\sqrt{2}}(\left\vert \uparrow \right\rangle _{A}\left\vert \downarrow
\right\rangle _{B}-\left\vert \downarrow \right\rangle _{A}\left\vert
\uparrow \right\rangle _{B})$ & $\left\vert \Psi \right\rangle _{A^{^{\prime
}}B^{^{\prime }}}^{-}=\frac{1}{\sqrt{2}}(\left\vert \downarrow \right\rangle
_{A^{^{\prime }}}\left\vert \uparrow \right\rangle _{B^{^{\prime
}}}-\left\vert \uparrow \right\rangle _{A^{^{\prime }}}\left\vert \downarrow
\right\rangle _{B^{^{\prime }}})$ \\ 
$\ \ P_{1}=1;$ $P_{2}=1$ & $\left\vert \Phi \right\rangle _{AB}^{+}=\frac{1}{%
\sqrt{2}}(\left\vert \uparrow \right\rangle _{A}\left\vert \uparrow
\right\rangle _{B}+\left\vert \downarrow \right\rangle _{A}\left\vert
\downarrow \right\rangle _{B})$ & $\left\vert \Phi \right\rangle
_{A^{^{\prime }}B^{^{\prime }}}^{+}=\frac{1}{\sqrt{2}}(\left\vert \downarrow
\right\rangle _{A^{^{\prime }}}\left\vert \downarrow \right\rangle
_{B^{^{\prime }}}+\left\vert \uparrow \right\rangle _{A^{^{\prime
}}}\left\vert \uparrow \right\rangle _{B^{^{\prime }}})$ \\ 
$\ \ P_{1}=1;$ $P_{2}=0$ & $\left\vert \Phi \right\rangle _{AB}^{-}=\frac{1}{%
\sqrt{2}}(\left\vert \uparrow \right\rangle _{A}\left\vert \uparrow
\right\rangle _{B}-\left\vert \downarrow \right\rangle _{A}\left\vert
\downarrow \right\rangle _{B})$ & $\left\vert \Phi \right\rangle
_{A^{^{\prime }}B^{^{\prime }}}^{-}=\frac{1}{\sqrt{2}}(\left\vert \downarrow
\right\rangle _{A^{^{\prime }}}\left\vert \downarrow \right\rangle
_{B^{^{\prime }}}-\left\vert \uparrow \right\rangle _{A^{^{\prime
}}}\left\vert \uparrow \right\rangle _{B^{^{\prime }}})$ \\ \hline\hline
\end{tabular}

\newpage \FRAME{ftbpF}{5.2797in}{3.9185in}{0pt}{}{}{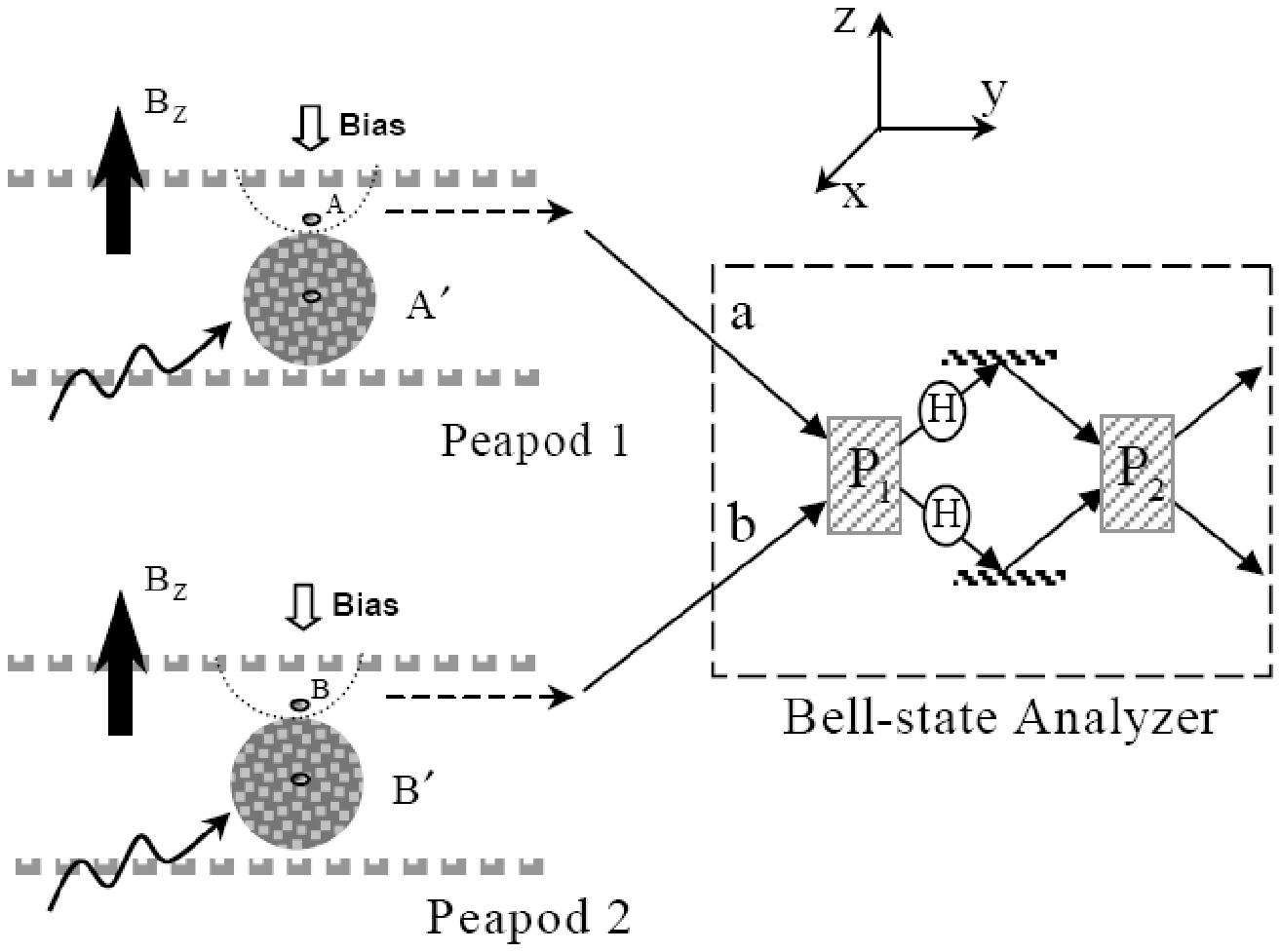}{\special{language
"Scientific Word";type "GRAPHIC";maintain-aspect-ratio TRUE;display
"USEDEF";valid_file "F";width 5.2797in;height 3.9185in;depth
0pt;original-width 5.2235in;original-height 3.87in;cropleft "0";croptop
"1";cropright "1";cropbottom "0";filename '1.eps';file-properties "XNPEU";}}

\FRAME{ftbpF}{4.3267in}{2.2805in}{0pt}{}{}{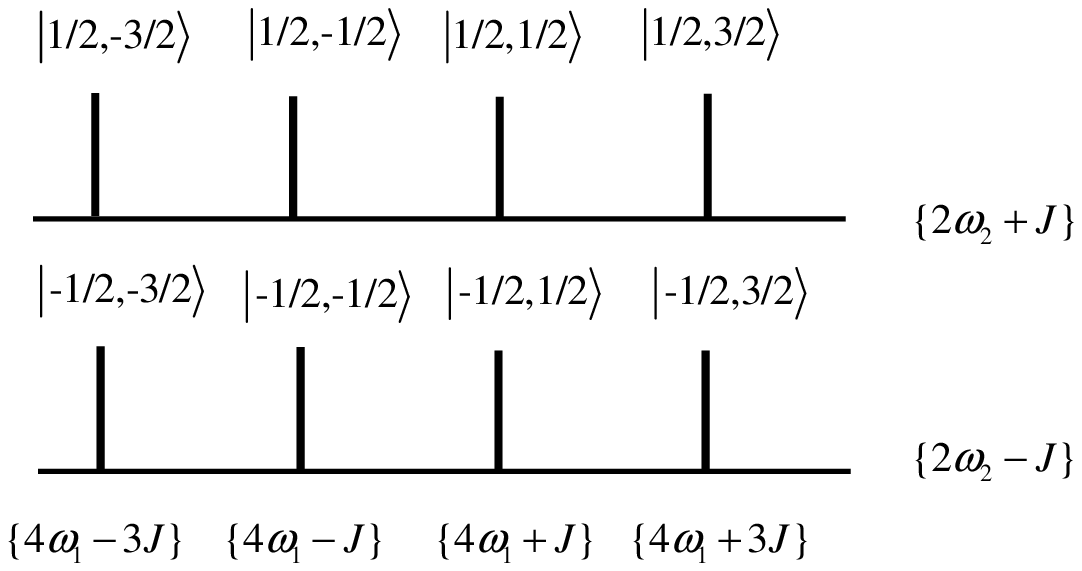}{\special{language
"Scientific Word";type "GRAPHIC";maintain-aspect-ratio TRUE;display
"USEDEF";valid_file "F";width 4.3267in;height 2.2805in;depth
0pt;original-width 4.2756in;original-height 2.2407in;cropleft "0";croptop
"1";cropright "1";cropbottom "0";filename '2.eps';file-properties "XNPEU";}}

\newpage

\FRAME{ftbpF}{5.2745in}{3.9133in}{0pt}{}{}{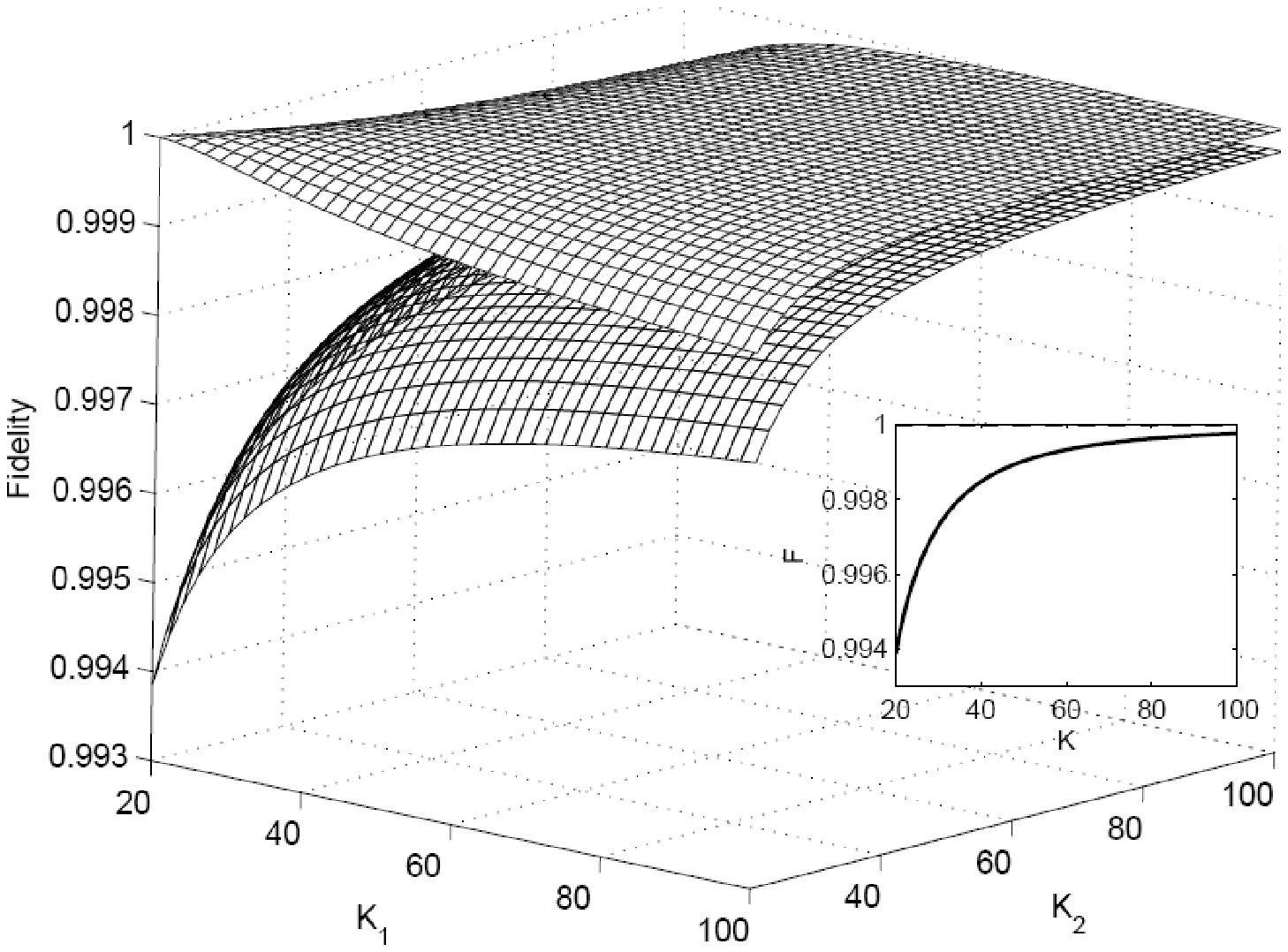}{\special{language
"Scientific Word";type "GRAPHIC";maintain-aspect-ratio TRUE;display
"USEDEF";valid_file "F";width 5.2745in;height 3.9133in;depth
0pt;original-width 6.6668in;original-height 4.9372in;cropleft "0";croptop
"1";cropright "1";cropbottom "0";filename '3.eps';file-properties "XNPEU";}}%
\FRAME{ftbpF}{5.2122in}{3.9115in}{0in}{}{}{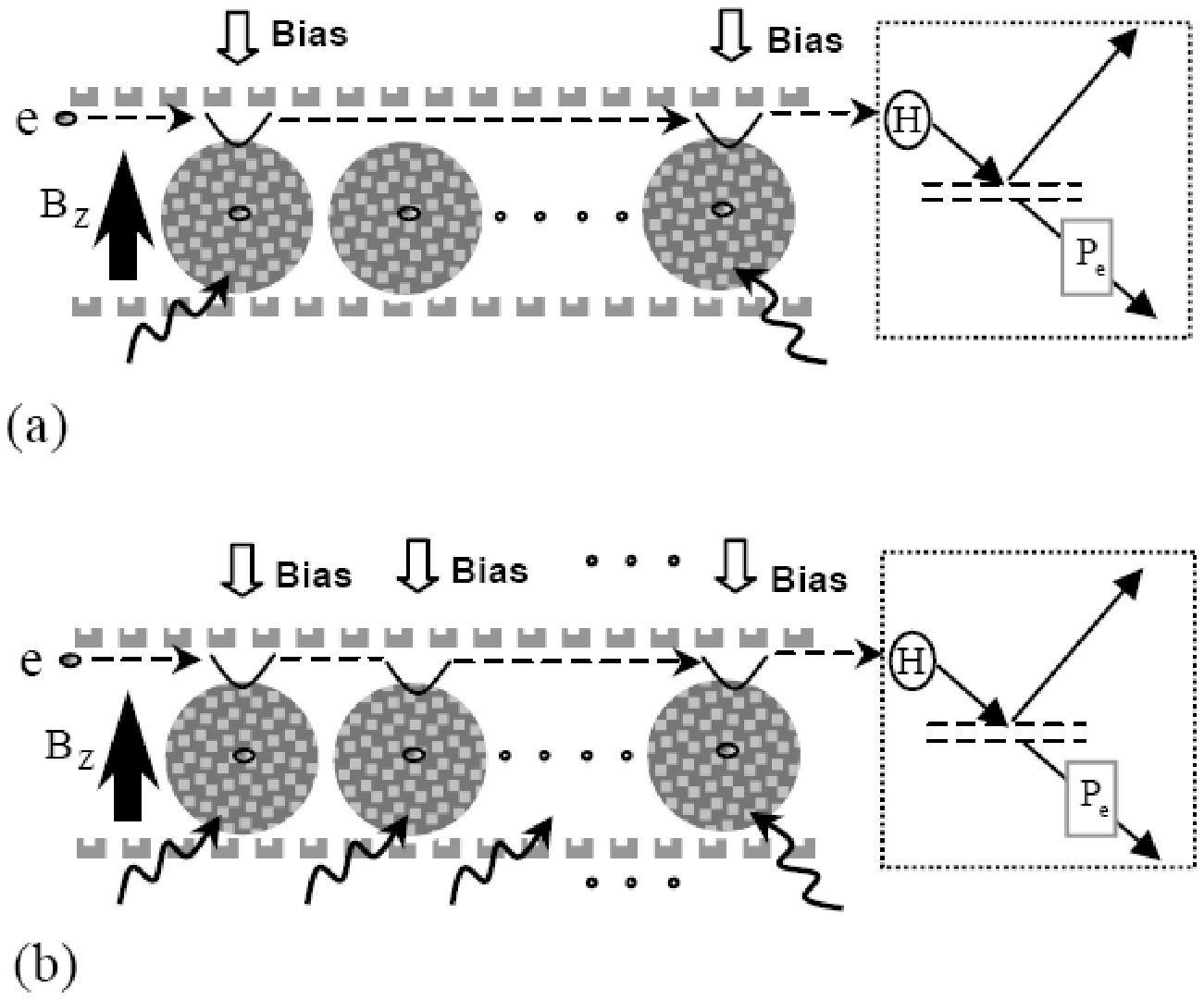}{\special{language
"Scientific Word";type "GRAPHIC";maintain-aspect-ratio TRUE;display
"USEDEF";valid_file "F";width 5.2122in;height 3.9115in;depth
0in;original-width 7.1105in;original-height 5.3255in;cropleft "0";croptop
"1";cropright "1";cropbottom "0";filename '4.eps';file-properties "XNPEU";}}

\ \ \ \ \ \ \ \ \ \ \ \ \ \ \ 


\begin{thebibliography}{99}
\bibitem{explain3} This simplified treatment is correct only under the
assumption that the operation after the $CNOT$ gating is very fast and
thereby the generated dephasing is negligible with respect to the phase
generated before the $CNOT$ gating. A more strict study with master equation
under the transport theory is underway.
\bibitem{explain} The electronic conductivity of carbon nanotube depends
heavily on the tube diameter and wrapping angle, with slight shift of which
the carbon nanotube could change from metallic state to semiconducting onr.
So we may cite experimental data from both the metallic case and the
semiconducting case.
\bibitem{explain2} Consider a line of fullerenes confined in each peapod and
the encapsulated electron spins have been entangled. Supposing that the two
endohedral fullerenes plotted in Fig. 1 are the end ones of the fullerene
lines, we would fuse the two lines of entangled electron spins in the
peapods by entangling the mobile qubits by a Bell-state analyzer.

{\Large Captions of Figures}
\end{thebibliography}
\end{document}